\documentstyle[aasms4]{article}
\makeatletter
  
  \@addtoreset{equation}{section}
\makeatother

\begin{document}

\title{RELATIVISTIC CORRECTIONS TO THE SUNYAEV-ZEL'DOVICH EFFECT FOR CLUSTERS OF GALAXIES. III. POLARIZATION EFFECT}

\author{NAOKI ITOH\altaffilmark{1}}

\affil{Department of Physics, Sophia University, 7-1 Kioi-cho, Chiyoda-ku, Tokyo, 102-8554, Japan}

\author{SATOSHI NOZAWA\altaffilmark{2}}

\affil{Josai Junior College for Women, 1-1 Keyakidai, Sakado-shi, Saitama, 350-0295, Japan}

\centerline{AND}

\author{YASUHARU KOHYAMA\altaffilmark{3}}

\affil{Fuji Research Institute Corporation, 2-3 Kanda-Nishiki-cho, Chiyoda-ku, Tokyo, 101-8443, Japan}

\altaffiltext{1}{n\_itoh@hoffman.cc.sophia.ac.jp}
\altaffiltext{2}{snozawa@galaxy.josai.ac.jp}
\altaffiltext{3}{kohyama@star.fuji-ric.co.jp}

\begin{abstract}

  We extend the formalism of the relativistic thermal and kinematic Sunyaev-Zel'dovich effects and include the polarization of the cosmic microwave background photons.  We consider the situation that a cluster of galaxies is moving with a velocity $\vec{\beta} \equiv \vec{v}/c$ with respect to the cosmic microwave background radiation.  In the present formalism, polarization of the scattered cosmic microwave background radiation caused by the proper motion of a cluster of galaxies is naturally derived as a special case of the kinematic Sunyaev-Zel'dovich effect.  The relativistic corrections are included also in a natural way.  Our results are in complete agreement with the recent results of the relativistic corrections obtained by Challinor, Ford \& Lasenby with an entirely different method as well as the nonrelativistic limit obtained by Sunyaev \& Zel'dovich.  The relativistic correction becomes to be significant in the Wien region.

\end{abstract}

\keywords{cosmic microwave background --- cosmology: theory --- galaxies: clusters: general ---  radiation mechanisms: thermal --- polarization --- relativity}

\section{INTRODUCTION}

  The present authors (Itoh, Kohyama, \& Nozawa 1998; Nozawa, Itoh, \& Kohyama 1998) have recently given accurate relativistic corrections to the thermal and kinematic Sunyaev-Zel'dovich effects (Zel'dovich \& Sunyaev 1969; Sunyaev \& Zel'dovich 1972,1980a, 1980b, 1981).  Their method was based on the kinetic equation for the photon distribution function using a relativistically covariant formalism (Berestetskii, Lifshitz, \& Pitaevskii 1982; Buchler \& Yueh 1976).  By using the generalized Kompaneets equation (Kompaneets 1957; Weymann 1965), they have derived analytic expressions for the thermal and kinematic Sunyaev-Zel'dovich effects as a power series of $\theta_{e}$ = $k_{B}T_{e}/mc^{2}$ and $\beta$, where $T_{e}$ is the electron temperature and $\beta$ is the peculiar velocity of the cluster divided by the velocity of light.  It has been shown by several groups (Stebbins 1997; Challinor \& Lasenby 1998; Itoh, Kohyama, \& Nozawa 1998) that the results for the thermal Sunyaev-Zel'dovich effect obtained by the power series expansion agree with the previous numerical calculations by Rephaeli (1995) and by Rephaeli \& Yankovitch (1997), thereby proving the validity of their method.  In particular, the convergence of the power series expansion has been carefully studied in the paper by Itoh, Kohyama, \& Nozawa (1998), where the analytic expressions up to $O(\theta_{e}^{5})$ have been derived and the results have been compared with those of the direct numerical integration of the Boltzmann equation.  It has been shown that the power series expansion approximation is sufficiently accurate for the region $k_{B}T_{e} \leq 15$keV by taking into account up to $O(\theta_{e}^{5})$ contributions.

  In the paper by Nozawa, Itoh, \& Kohyama (1998), a formalism for the kinematic Sunyaev-Zel'dovich effect for the clusters of galaxies with a peculiar velocity $\beta$ has been derived by applying the Lorentz boost to the standard formalism of the extended Kompaneets equation.  With the power series expansion approximation in terms of the electron temperature $\theta_{e}$ and the peculiar velocity $\beta$, an analytic expression for the kinematic Sunyaev-Zel'dovich effect which includes the relativistic corrections of up to $O(\beta \theta_{e}^{2})$ and $O(\beta^{2} \theta_{e})$ has been derived.  It has been found that the relativistic correction is significant.  Similar works have been reported by Sazonov \& Sunyaev (1998), and also by Challinor \& Lasenby (1999).  They are in essential agreement with the works of the present authors.

  In the present paper we address ourselves to the problem of the polarization of the cosmic microwave background radiation (CMBR) caused by the proper motion of the cluster of galaxies.  This problem has been studied by Sunyaev \& Zel'dovich (1980b) in the nonrelativistic limit.  The polarization Sunyaev-Zel'dovich effect enables the measurement of the tangential velocities of the clusters of galaxies, thus contributing to the measurement of the radial velocities complemented by the measurement of the velocities along the line of sight which is achieved by means of the observation of the kinematic Sunyaev-Zel'dovich effect.  There exist,
however, competing polarization effects connected with the finite depth of the intracluster plasma.  These effects have been discussed by Sunyaev \& Zel'dovich (1980b), and more recently by Sazonov \& Sunyaev (1999).
In the present paper we solve this problem by extending our already-established formalism to the case of polarized photons.  With the present formalism the effect of the polarization of the CMBR can be derived on the same footing as the calculation of the thermal and kinematic Sunyaev-Zel'dovich effects.  We thereby take into account the relativistic corrections accurately.  

  After the submittal of the original manuscript of the present paper to The Astrophysical Journal, we became aware of the similar works by Audit \& Simmons (1998) as well as by Hansen \& Lilje (1999).  More recently the papers by Sazonov \& Sunyaev (1999) and by Challinor, Ford \& Lasenby (1999) have been submitted on the similar subject (the polarization of the CMBR).  Through the communication with these authors, we have benefited a lot and have revised the present paper.  In particular, Dr. Challinor has kindly pointed out an error in the original manuscript of the present paper.  In the present revised manuscript, we have fully taken into account their comments and have corrected the error in the original manuscript.  With the present formalism we have confirmed the recent result obtained by Challinor, Ford \& Lasenby (1999) with an entirely different method as well as the nonrelativistic limit obtained by Sunyaev \& Zel'dovich (1980b).

  The present paper is organized as follows: the Kompaneets equation for polarized photons will be derived in $\S$ 2.  Results of the calculation will be presented in $\S$ 3.  Concluding remarks will be given in $\S$ 4.

\section{LORENTZ BOOSTED KOMPANEETS EQUATION}

  In the present section we will extend the Kompaneets equation studied by Nozawa, Itoh, \& Kohyama (1998) to the case of polarized photons.  We will consider a cluster of galaxies moving with a peculiar velocity with respect to the cosmic microwave background radiation (CMBR).  We will formulate the kinetic equation for the photon distribution function using a relativistically covariant formalism (Berestetskii, Lifshitz, \& Pitaevskii 1982; Buchler \& Yueh 1976).  As a reference system, we choose the system which is fixed to CMBR.  The $z$-axis is fixed to the line connecting the cluster of galaxies (CG) and the observer. (We assume that the observer is fixed to the CMBR frame.)  We fix the positive direction of the $z$-axis as the direction of the propagation of a photon from the cluster to the observer.  In this reference system, the center of mass of CG is moving with a peculiar velocity $\vec{\beta} (\equiv \vec{v}/c$) with respect to the CMBR.  For simplicity, we choose the direction of the velocity in the $x$-$z$ plane, i.e. $\vec{\beta} = (\beta_{x}, 0, \beta_{z})$.

  In order to describe the Compton scattering of polarized photons by unpolarized electrons, we derive the Stokes parameters using the polarization density matrix (Berestetskii, Lifshitz, \& Pitaevskii (1982)).  We here emphasize the importance of using the relativistically covariant formalism.  Sunyaev \& Zel'dovich (1980b) have shown that the degree of the polarization of the cosmic microwave background photon caused by the proper motion of the cluster of galaxies is of order $\beta^{2}$.  On the other hand, Nozawa, Itoh, \& Kohyama (1998) have derived the relativistic corrections to the kinematical Sunyaev-Zel'dovich effect which are of order $\beta^{2}$.  Therefore, one has to be extremely careful to take into account all the relevant terms of order $\beta^{2}$ when one calculates the polarization Sunyaev-Zel'dovich effect.  This is the reason why we adopt the relativistically covariant formalism in this paper.

In the CMBR frame, the time evolution of the photon distribution $n( \omega)$ is written as follows.
\begin{eqnarray}
\frac{\partial n(\omega) }{\partial t}  & = & - 2 \int \frac{d^{3}p}{(2\pi)^{3}} d^{3}p^{\prime} d^{3}k^{\prime} \, 
\frac{(e^{2}/4\pi)^{2}  \, \delta^{4}(p+k-p^{\prime}-k^{\prime})}{2 \omega \omega^{\prime} E E^{\prime}} \, \hspace{5.0cm}  \nonumber  \\
& \times & \left\{ \, X_{k + p \rightarrow k^{\prime} + p^{\prime}} \, 
n(\omega) [1 + n(\omega^{\prime})] f(E) \, - \, X_{k^{\prime} + p^{\prime} \rightarrow k + p} \, n(\omega^{\prime})[1 + n(\omega)] f(E^{\prime}) \,  \right\}   \,  ,
\end{eqnarray}
\noindent
where $k, k^{\prime}$ are the four-momenta of photons and $p, p^{\prime}$ are the four-momenta of electrons, respectively.  In equation (2.1) the first term corresponds to the Compton scattering $k + p \rightarrow k^{\prime} + p^{\prime}$.  The explicit form is given as follows.
\begin{eqnarray}
X_{k + p \rightarrow k^{\prime} + p^{\prime}} & = & \overline{X} \, \left[ \, 1 \, + \, \vec{\eta} \cdot \vec{\zeta}^{(i)} \, \right]  \, ,  \hspace{7.0cm}
\end{eqnarray}
where
\begin{eqnarray}
\overline{X} & = & 4 \left\{  \left( \frac{1}{x} - \frac{1}{y} \right)^{2} 
\, + \, \left( \frac{1}{x} - \frac{1}{y} \right) \, + \, \frac{1}{4} \left( \frac{y}{x} + \frac{x}{y} \right)  \right\} \,  ,  \\
\zeta_{1}^{(i)} & = & \frac{1}{ \overline{X}} \, \frac{8}{m^{2}} \left\{ - \left( \frac{1}{x} - \frac{1}{y} \right)^{2} e^{(1)} \cdot p \, e^{(2)} \cdot p \, - \, \left( \frac{1}{x} - \frac{1}{y} \right) \frac{1}{y} \, e^{(1)} \cdot p \, e^{(2)} \cdot k^{\prime} \right\}
  \, ,  \\
\zeta_{2}^{(i)} & = & 0 \, ,  \\
\zeta_{3}^{(i)} & = & \frac{1}{ \overline{X}} \, \frac{4}{m^{2}} \, \left\{ - \left( \frac{1}{x} - \frac{1}{y} \right)^{2} \, \left[ \left( e^{(1)} \cdot p \right)^{2} \, - \, \left( e^{(2)} \cdot p \right)^{2} \right] \,  \right.  \nonumber  \\
&   & \left. \hspace{1.7cm} \, + \, 2 \left( \frac{1}{x} - \frac{1}{y} \right) \frac{1}{y} \, e^{(2)} \cdot p \, e^{(2)} \cdot k^{\prime} \, 
+ \, \frac{1}{y^{2}} \, \left( e^{(2)} \cdot k^{\prime} \right)^{2}  \, \right\}   \, , \\
  \nonumber  \\
x & \equiv & \left(s \, - \, m^{2} \right)/m^{2}  \,  ,  \\
y & \equiv & \left(m^{2} \, - \, u \right)/m^{2}  \,  ,  \\
s & = & \left( p + k \right)^{2} \, = \, \left( p^{\prime} + k^{\prime} \right)^{2}  \,  ,  \\
u & = & \left( p - k^{\prime} \right)^{2} \, = \, \left( p^{\prime} - k \right)^{2}  \,   .
\end{eqnarray}
In equation (2.2), $\vec{\eta} = ( \eta_{1}, \eta_{2}, \eta_{3} )$ is the Stokes parameter which corresponds to the selection of the polarization of the photon of momentum $k$ by the detector (Berestetskii, Lifshitz, \& Pitaevskii (1982)).  In equations (2.4)--(2.6), $e^{(1)}$ and $e^{(2)}$ are the polarization vectors for photons of the momentum $k$.  We choose the explicit forms as follows:
\begin{eqnarray}
k & = & ( \omega, \vec{k}) \, , \, \hspace{1.2cm} \, \vec{k} \, = \, (0, 0, \omega)  \, ,  \\
e^{(1)} & = & ( 0, \vec{e}^{ \, (1)} ) \, , \, \hspace{0.5cm} \, \vec{e}^{ \, (1)} \, = \, \frac{ \vec{k} \times \vec{k}^{\prime} }{ \mid \vec{k} \times \vec{k}^{\prime} \mid }  \, ,  \\
e^{(2)} & = & ( 0, \vec{e}^{ \, (2)} ) \, , \, \hspace{0.5cm} \, \vec{e}^{ \, (2)} \, = \, \frac{ \vec{k} \times ( \vec{k} \times \vec{k}^{\prime} ) }{ \mid \vec{k} \times ( \vec{k} \times \vec{k}^{\prime} ) \mid }  \, .
\end{eqnarray}

  Similarly, the second term in equation (2.1) corresponds to the Compton scattering $k^{\prime} + p^{\prime} \rightarrow k + p$.  The explicit form is as follows:
\begin{eqnarray}
X_{k^{\prime} + p^{\prime} \rightarrow k + p} & = & \overline{X} \, 
\left[ \, 1 \, + \, \vec{\eta} \cdot \vec{\zeta}^{(f)} \, \right]  \, , \hspace{7.0cm}
\end{eqnarray}
where
\begin{eqnarray}
\zeta_{1}^{(f)} & = & \frac{1}{\overline{X}} \, \frac{8}{m^{2}} \left\{ - \left( \frac{1}{x} - \frac{1}{y} \right)^{2} e^{(1)} \cdot p^{\prime} \, e^{(2)} \cdot p^{\prime} \, - \, \left( \frac{1}{x} - \frac{1}{y} \right) \frac{1}{x} \, e^{(1)} \cdot p^{\prime} \, e^{(2)} \cdot k^{\prime} \right\}   \, ,  \\
\zeta_{2}^{(f)} & = & 0 \, ,  \\
\zeta_{3}^{(f)} & = & \frac{1}{ \overline{X}} \, \frac{4}{m^{2}} \, \left\{ - \left( \frac{1}{x} - \frac{1}{y} \right)^{2} \, \left[ \left( e^{(1)} \cdot p^{\prime} \right)^{2} \, - \, \left( e^{(2)} \cdot p^{\prime} \right)^{2} \right] \,  \right.  \nonumber  \\
&   & \left. \hspace{1.7cm} \, + \, 2 \left( \frac{1}{x} - \frac{1}{y} \right) \frac{1}{x} \, e^{(2)} \cdot p^{\prime} \, e^{(2)} \cdot k^{\prime} \, 
+ \, \frac{1}{x^{2}} \, \left( e^{(2)} \cdot k^{\prime} \right)^{2}  \, \right\}   \, .
\end{eqnarray}

Inserting equations (2.2) and (2.14) into equation (2.1), we obtain
\begin{eqnarray}
\frac{\partial n(\omega) }{\partial t}  & = & - 2 \int \frac{d^{3}p}{(2\pi)^{3}} d^{3}p^{\prime} d^{3}k^{\prime} \, W \,
\left\{ n(\omega) [1 + n(\omega^{\prime})] f(E) \, - \, n(\omega^{\prime})[1 + n(\omega)] f(E^{\prime}) \,  \right\}   \, \nonumber  \\
& & - 2 \int \frac{d^{3}p}{(2\pi)^{3}} d^{3}p^{\prime} d^{3}k^{\prime} \, W \, \vec{\eta} \cdot  \left\{ \, \, \vec{\zeta}^{(i)} n(\omega) [1 + n(\omega^{\prime})] f(E) \, \right.  \nonumber  \\
&  & \left. \hspace{4.5cm} - \, \vec{\zeta}^{(f)} \, n(\omega^{\prime})[1 + n(\omega)] f(E^{\prime}) \,  \right\}   \,  ,
\end{eqnarray}
where
\begin{eqnarray}
W & = & \frac{(e^{2}/4\pi)^{2}  \, \delta^{4}( p+k-p^{\prime}-k^{\prime} )}{2 \omega \omega^{\prime} E E^{\prime}} \,  \overline{X}  \,
\end{eqnarray}
is the transition probability corresponding to the Compton scattering of an unpolarized photon by an unpolarized electron.  In equation (2.18) the first line corresponds to the unpolarized case, which has been studied in Itoh, Kohyama, \& Nozawa (1998) and in Nozawa, Itoh, \& Kohyama (1998).  The second and third lines correspond to the polarized photon contribution.

Equation (2.18) is further simplified.  With the help of the following relations
\begin{eqnarray}
e^{(1)} \cdot p^{\prime} & = & e^{(1)} \cdot (p + k - k^{\prime})  \, = \, e^{(1)} \cdot p  \, , \\
e^{(2)} \cdot p^{\prime} & = & e^{(2)} \cdot (p + k - k^{\prime})  \, = \, e^{(2)} \cdot p \, - \, e^{(2)} \cdot k^{\prime}   \, ,
\end{eqnarray}
it is straightforward to show that
\begin{eqnarray}
\vec{\zeta} & \equiv & \vec{\zeta}^{(i)} \, = \, \vec{\zeta}^{(f)} \,  .
\end{eqnarray}
Therefore we have
\begin{eqnarray}
\frac{\partial n(\omega) }{\partial t}  & = & - 2 \int \frac{d^{3}p}{(2\pi)^{3}} d^{3}p^{\prime} d^{3}k^{\prime} \, W \,
\left\{ n(\omega) [1 + n(\omega^{\prime})] f(E) \, - \, n(\omega^{\prime})[1 + n(\omega)] f(E^{\prime}) \,  \right\}   \, \nonumber  \\
& & - 2 \int \frac{d^{3}p}{(2\pi)^{3}} d^{3}p^{\prime} d^{3}k^{\prime} \, W \, \vec{\eta} \cdot \vec{\zeta} \, \left\{ \,  n(\omega) [1 + n(\omega^{\prime})] f(E) \, \right.  \nonumber  \\
&  &  \left.  \hspace{4.8cm} - \, n(\omega^{\prime})[1 + n(\omega)] f(E^{\prime}) \,  \right\}   \,   .
\end{eqnarray}
The vector $\vec{\zeta}$ represents the Stokes parameter of the photon $k$ which is polarized due to the Compton scattering (Berestetskii, Lifshitz, \& Pitaevskii (1982)).

  Next we transform the Stokes parameter $\vec{\zeta}$=$(\zeta_{1},\zeta_{2},\zeta_{3})$ to the Stokes parameter $\vec{\xi}$=$(\xi_{1},\xi_{2},\xi_{3})$ defined with respect to the coordinate system $xyz$ fixed to the CMBR.  According to Landau \& Lifshitz (1975), we obtain
\begin{eqnarray}
\xi_{1} & = & \zeta_{1} \, {\rm cos} \, 2 \left[ - \left( \phi_{\gamma}^{\prime} + \frac{ \pi}{2} \right) \right] \, - \, \zeta_{3} \, {\rm sin} \, 2 \left[ - \left( \phi_{\gamma}^{\prime} + \frac{ \pi}{2} \right) \right]  \, ,  \\
\xi_{2} & = & \zeta_{2}  \,  ,   \\
\xi_{3} & = & \zeta_{1} \, {\rm sin} \, 2 \left[ - \left( \phi_{\gamma}^{\prime} + \frac{ \pi}{2} \right) \right] \, + \, \zeta_{3} \, {\rm cos} \, 2 \left[ - \left( \phi_{\gamma}^{\prime} + \frac{ \pi}{2} \right) \right]  \, , 
\end{eqnarray}
where $\phi_{\gamma}^{\prime}$ is the azimuthal angle of the vector $\vec{k}^{\prime}$ with respect to the $x$-$z$ plane.  Since $\zeta_{2}=0$, we obtain $\xi_{2}=0$.  Therefore, we have 
\begin{eqnarray}
 \left( \begin{array}{ll} \xi_{1}  \\   \xi_{3} \end{array}  \right)
 & = & \left( \begin{array}{cccc}  - \, {\rm cos} \, 2 \phi_{\gamma}^{\prime}  & - \, {\rm sin} \, 2 \phi_{\gamma}^{\prime}  \\
  \, \, \, \, \, {\rm sin} \, 2 \phi_{\gamma}^{\prime} & - \, {\rm cos} \, 2 \phi_{\gamma}^{\prime}  \end{array} \right)  \left( \begin{array}{ll} \zeta_{1}  \\   \zeta_{3} \end{array}  \right) \, ,
\end{eqnarray}
where $\zeta_{1}=\zeta_{1}^{(f)}$ and $\zeta_{3}=\zeta_{3}^{(f)}$ are given by equations (2.15) and (2.17).

  Thus, the time evolution of the Stokes parameter corresponding to the photon polarization in the CMBR frame is written as follows (Acquista \& Anderson 1974; Nagirner 1994):
\begin{eqnarray}
\frac{\partial}{\partial t} 
\left[ n(\omega) \left( \begin{array}{ll} \xi_{1} \\  \xi_{3} \end{array}    \right) \right]  & = &  - 2 \int \frac{d^{3}p}{(2\pi)^{3}} d^{3}p^{\prime} d^{3}k^{\prime} \, W \, \left( \begin{array}{cccc}  - {\rm cos} \, 2 \phi_{\gamma}^{\prime}  & -  {\rm sin} \, 2 \phi_{\gamma}^{\prime}  \\
  \, \, \, \, \, {\rm sin} \, 2 \phi_{\gamma}^{\prime} & - {\rm cos} \, 2 \phi_{\gamma}^{\prime}  \end{array} \right) \, \left( \begin{array}{ll} \zeta_{1}  \\   \zeta_{3} \end{array}  \right)  \,  \nonumber \\
  & \times &  \left\{ \, n(\omega)[1 + n(\omega^{\prime})] f(E)  \, -  \, n(\omega^{\prime})[1 + n(\omega)] f(E^{\prime}) \, \right\} \,  .
\end{eqnarray} 
  The electron Fermi distribution functions in the initial and final states are defined in the CG frame.  They are related to the electron Fermi distribution functions in the CMBR frame as follows (Landau \& Lifshitz 1975):
\begin{eqnarray}
f(E) & = &  f_{C}(E_{C})  \, , \\
f(E^{\prime}) & = &  f_{C}(E_{C}^{\prime})  \,  ,  \\
E_{C} & = & \gamma \, \left(E - \vec{\beta} \cdot  \vec{p} \right) \, ,   \\
E_{C}^{\prime} & = & \gamma \, \left(E^{\prime} - \vec{\beta} \cdot \vec{p}^{\prime} \right) \, ,  \\
\gamma & \equiv & \frac{1}{\sqrt{1 - \beta^{2}}}   \, ,
\end{eqnarray}
where the suffix $C$ denotes the CG frame.  We caution the reader that the electron distribution function is anisotropic in the CMBR frame for $\beta \neq 0$.  By ignoring the degeneracy effects, we have the relativistic Maxwellian distribution for electrons with temperature $T_{e}$ as follows:
\begin{eqnarray}
f_{C}(E_{C}) & = & \left[ e^{\left\{(E_{C} - m)-(\mu_{C} - m) \right\}/k_{B}T_{e}} \, + \, 1 \right]^{-1}  \nonumber \\
& \approx & e^{-\left\{(E_{C}-m)-(\mu_{C} - m)\right\}/k_{B}T_{e}} \, ,
\end{eqnarray}
where $(\mu_{C} - m)$ is the non-relativistic chemical potential of the electron measured in the CG frame.  We now introduce the quantities
\begin{eqnarray}
x &  \equiv &  \frac{\omega}{k_{B}T_{e}}  \, ,  \\
\Delta x &  \equiv &  \frac{\omega^{\prime} - \omega}{k_{B}T_{e}}  \, .
\end{eqnarray}
Substituting equations (2.29) -- (2.36) into equation (2.28), we obtain
\begin{eqnarray}
\frac{\partial}{\partial t}
\left[ n(\omega) \left( \begin{array}{ll} \xi_{1} \\  \xi_{3} \end{array}    \right) \right]
 & = & -2 \int \frac{d^{3}p}{(2\pi)^{3}} d^{3}p^{\prime} d^{3}k^{\prime} \, W \,  
 \left( \begin{array}{cccc}  - {\rm cos} \, 2 \phi_{\gamma}^{\prime}  & -  {\rm sin} \, 2 \phi_{\gamma}^{\prime}  \\
  \, \, \, \, \, {\rm sin} \, 2 \phi_{\gamma}^{\prime} & - {\rm cos} \, 2 \phi_{\gamma}^{\prime}  \end{array} \right)  \left( \begin{array}{ll} \zeta_{1}  \\   \zeta_{3} \end{array}  \right)  \,   \nonumber  \\
 & \times & \, f_{C}(E_{C}) \, \left[ \, \left\{ \, 1 + n(\omega^{\prime}) \, \right\} n(\omega) \, \right.   \nonumber  \\
&  & \left. \hspace{1.5cm} - \,  \left\{ \, 1 + n(\omega) \, \right\} n(\omega^{\prime}) \, {\rm e}^{ \Delta x \gamma (1 - \vec{\beta} \cdot \hat{k}^{\prime} ) } \, {\rm e}^{ x \gamma \vec{\beta} \cdot ( \hat{k} - \hat{k}^{\prime} ) } \right] \, ,
\end{eqnarray}
where $\hat{k}$ and $\hat{k}^{\prime}$ are the unit vectors in the directions of $\vec{k}$ and $\vec{k}^{\prime}$, respectively.  Equation 
(2.37) is our basic equation.

\section{RESULTS OF THE CALCULATION}

  We expand the right-hand-side of equation (2.37) in powers of $\Delta x$ by assuming $\Delta x \ll 1$ as has been done by Nozawa, Itoh, \& Kohyama (1998):
\begin{eqnarray}
\frac{ \partial}{ \partial t} \left[ n(\omega) \left( \begin{array}{ll} \xi_{1} \\  \xi_{3} \end{array}    \right) \right] & = & 
 2 \left[ \frac{ \partial n}{ \partial x} \, \vec{I}_{1,0} + n(1+n) \, \vec{I}_{1,1} \right]
  \nonumber  \\
& + & 2 \left[ \frac{ \partial^{2} n}{ \partial x^{2}} \, \vec{I}_{2,0}
+ 2(1+n) \frac{ \partial n}{ \partial x} \, \vec{I}_{2,1} + n(1+n)  \, \vec{I}_{2,2} \right]
  \nonumber  \\
& + & 2 \left[\frac{ \partial^{3} n}{ \partial x^{3}} \, \vec{I}_{3,0}
+ 3(1+n) \frac{ \partial^{2} n}{ \partial x^{2}} \, \vec{I}_{3,1}
+ 3(1+n) \frac{ \partial n}{ \partial x} \, \vec{I}_{3,2} + n(1+n) \, \vec{I}_{3,3} \right]
  \nonumber \\
& + & \cdot \cdot \cdot  \,  \nonumber \\
& + & 2 \, n \, \left[ (1 + n) \vec{J}_{0} + \frac{ \partial n}{ \partial x } \,  \vec{J}_{1} + \frac{ \partial^{2} n}{ \partial x^{2}} \, \vec{J}_{2} + \frac{ \partial^{3} n}{ \partial x^{3}} \, \vec{J}_{3} + \cdot \cdot \cdot  \, \, \,  \right] \, \,  ,
\end{eqnarray}
where
\begin{eqnarray}
\vec{I}_{k, \ell} & \equiv & \frac{1}{k !} \int \frac{d^{3}p}{(2\pi)^{3}} d^{3}p^{\prime} d^{3}k^{\prime} \, W \,  \left( \begin{array}{cccc}  - {\rm cos} \, 2 \phi_{\gamma}^{\prime}  & -  {\rm sin} \, 2 \phi_{\gamma}^{\prime}  \\
  \, \, \, \, \, {\rm sin} \, 2 \phi_{\gamma}^{\prime} & - {\rm cos} \, 2 \phi_{\gamma}^{\prime}  \end{array} \right)  \left( \begin{array}{ll} \zeta_{1} \\   \zeta_{3} \end{array}  \right)  \, \nonumber  \\
&  &  \hspace{0.8cm} \times \, f_{C}(E_{C}) \, (\Delta x)^{k}  \,  {\rm e}^{ x \gamma \vec{\beta} \cdot ( \hat{k} - \hat{k}^{\prime} ) }  
\gamma^{ \ell} \left( 1 - \vec{\beta} \cdot \hat{k}^{\prime} 
\right)^{ \ell}  \, , \\
  \nonumber \\
\vec{J}_{k} & \equiv & - \frac{1}{k !} \int \frac{d^{3}p}{(2\pi)^{3}} d^{3}p^{\prime} d^{3}k^{\prime} \, W \,  \left( \begin{array}{cccc}  - {\rm cos} \, 2 \phi_{\gamma}^{\prime}  & -  {\rm sin} \, 2 \phi_{\gamma}^{\prime}  \\
  \, \, \, \, \, {\rm sin} \, 2 \phi_{\gamma}^{\prime} & - {\rm cos} \, 2 \phi_{\gamma}^{\prime}  \end{array} \right)  \left( \begin{array}{ll} \zeta_{1} \\   \zeta_{3} \end{array}  \right)  \, \nonumber  \\
&  &  \hspace{0.8cm} \times \, f_{C}(E_{C}) \, (\Delta x)^{k}  \left( \, 1 \, - \, {\rm e}^{ x \gamma \vec{\beta} \cdot ( \hat{k} - \hat{k}^{\prime} ) }  \right)  \,  .
\end{eqnarray}
\noindent
For most of the clusters of galaxies, $\beta \ll 1$ is realized.  For example, $\beta \approx 1/300$ for a typical value of the peculiar velocity $v=1,000$km/s.  Therefore it should be sufficient to expand in powers of $\beta$ and retain up to $O(\beta^{2})$ contributions.  We assume the initial photon distribution of the CMBR to be Planckian with a temperature $T_{0}$:
\begin{equation}
n_{0} (X) \, = \, \frac{1}{e^{X} - 1} \, , 
\end{equation}
where
\begin{equation}
X \, \equiv \, \frac{\omega}{k_{B} T_{0}}  \, .
\end{equation}

  Assuming $T_{0}/T_{e} \ll 1$, we obtain
\begin{eqnarray}
P  \, \equiv  \, \frac{1}{n_{0}(X)} \, \Delta \, \left[ n(X) \left( \begin{array}{ll} \xi_{1} \\  \xi_{3} \end{array}  \right) \right] & = & \frac{y \, X e^{X}}{e^{X}-1} \,  \beta_{x}^{2} \left(  \, \, F_{0} \, + \, \theta_{e} F_{1}  \, + \, \theta_{e}^{2} F_{2} \, \, \right) \, \left( \begin{array}{ll}  0  \\  1 \end{array}  \right) \, ,  \hspace{1.0cm} \\
  \nonumber  \\
\beta_{x} & = & \beta \, {\rm sin} \, \theta_{\gamma}  \, ,  \\
y & \equiv & \sigma_{T} \int d \ell N_{e}  \, , \\
\theta_{e} & = & \frac{k_{B}T_{e}}{mc^{2}}  \, ,  \\
  \nonumber  \\
F_{0} & = &   \frac{1}{20} \tilde{X} \, , \\
  \nonumber  \\
F_{1} & = & \frac{3}{10}  \tilde{X} - \frac{2}{5} \left( \tilde{X}^{2} + \frac{1}{2} \tilde{S}^{2} \right) + \frac{1}{20} \left( \tilde{X}^{3} + 2 \tilde{X} \tilde{S}^{2} \right)   \,  ,  \\
  \nonumber  \\
F_{2} & = &  \frac{3}{4}  \tilde{X} - \frac{21}{5} \left( \tilde{X}^{2} + \frac{1}{2} \tilde{S}^{2} \right) + \frac{867}{280} \left( \tilde{X}^{3} + 2 \tilde{X} \tilde{S}^{2} \right) \,  \nonumber  \\
& - &  \frac{4}{7} \left( \tilde{X}^{4} + \frac{11}{2} \tilde{X}^{2} \tilde{S}^{2} + \tilde{S}^{4} \right) \nonumber  \\
 & + &  \frac{1}{35} \left( \tilde{X}^{5} + 13 \tilde{X}^{3} \tilde{S}^{2} + \frac{17}{2} \tilde{X} \tilde{S}^{4} \right)  \, ,   \\
  \,  \nonumber  \\
\tilde{X} & \equiv &  X \, {\rm coth} \left( \frac{X}{2} \right)  \, , \\
\tilde{S} & \equiv & \frac{X}{ \displaystyle{ {\rm sinh} \left( \frac{X}{2} \right)} }   \, , 
\end{eqnarray}
where $\theta_{\gamma}$ is the angle between the directions of the peculiar velocity ($\vec{\beta}$) and the photon momentum ($\vec{k}$) which is chosen as the $z$-direction.  $N_{e}$ is the electron number density in the CG frame.  The integral in equation (3.8) is over the photon path length in the cluster, and $\sigma_{T}$ is the Thomson cross section.  

  Equations (3.6)--(3.14) are in complete agreement with the recent result obtained by Challinor, Ford \& Lasenby (1999) with an entirely different method.  It should be emphasized that the effect of the polarization of the CMBR has been derived on the same footing as the calculation of the thermal and kinematic Sunyaev-Zel'dovich effects by using the present formalism.  It is also worthwhile to mention here that the polarization (equation (3.6)) is proportional to $\beta_{x}^{2}$, where $\beta_{x}$ is the {\it transverse component} of the peculiar velocity of CG.  In the present case the polarization of the CMBR is nonzero, because the electron distribution is anisotropic in the CMBR frame for the nonzero peculiar velocity.  It should be also noted that we have neglected higher order relativistic corrections such as $O(\theta_{e}^{3})$ terms in deriving equation (3.6).  This is because the observation of the higher order term in the polarization of the CMBR caused by the motion of the CG will not be feasible in the near future.  Therefore it should be sufficient to neglect higher order relativistic corrections.

  In the Rayleigh-Jeans limit $X \rightarrow 0$ with $\theta_{e}=0$, equation (3.6) gives the polarization degree
\begin{eqnarray}
P  & = &  \frac{1}{10} \, y \, \beta_{x}^{2}  \, \, ,
\end{eqnarray}
which agrees with the result of Sunyaev \& Zel'dovich (1980b).  Similar results have been reported in the more recent works: Audit \& Simmons (1998); Hansen \& Lilje (1999); Sazonov \& Sunyaev (1999); Challinor, Ford \& Lasenby (1999).

  In Figure 1 we show the graph of the function
\begin{equation}
\Xi \, \equiv \, \frac{1}{y \beta_{x}^{2}} \frac{ \Delta \left[ n(X) \xi_{3} \right] }{ n_{0}(X)} \, = \,  \frac{X e^{X}}{e^{X}-1} \, \left(  \, \, F_{0} \, + \, \theta_{e} F_{1}  \, + \, \theta_{e}^{2} F_{2}  \, \, \right) \,  .
\end{equation}
It is clear from Figure 1 that the value of the polarization function $\Xi$ is positive (linear polarization in the plane formed by the velocity vector of the cluster and the vector which connects the cluster and the observer) and small for the Rayleigh-Jeans region, however, it quickly becomes large for the Wien region, where the coefficient is larger than 10 at $X \geq 11$.   Finally we define the distribution of the spectral intensity of the polarized radiation as follows:
\begin{eqnarray}
\Delta I_{pol} & \equiv & X^{3} \Delta \left[ n(X)  \xi_{3} \right] \, = \,  y \beta_{x}^{2} \frac{X^{4} e^{X}}{\left(e^{X}-1 \right)^{2} } \, \left(  \, \, F_{0} \, + \, \theta_{e} F_{1}  \, + \, \theta_{e}^{2} F_{2} \, \, \right) \,  .
\end{eqnarray}
The graph of the function $\Delta I_{pol}/y$ is shown in Figure 2.  It has been found that the function has a maximum value of $11\times10^{-6}$ at $X \approx 5$ for $k_{B}T = 10$keV, $\beta_{x} = 1/300$.  It is clear that the relativistic correction becomes significant for large values of $X$, typically for $X>5$. 

\section{CONCLUDING REMARKS}

  We have calculated the Stokes parameter corresponding to the polarization of the CMBR caused by the proper motion of a cluster of galaxies.  The calculation is based on the covariant formalism.  With the present formalism the polarization of the CMBR is calculated as a natural extension of the kinematic Sunyaev-Zel'dovich effect calculated by Nozawa, Itoh, \& Kohyama (1998).  Thus the relativistic corrections have been derived from first principles.  We have confirmed the recent result of the relativistic corrections obtained by Challinor, Ford, \& Lasenby (1999) with an entirely different method as well as the nonrelativistic limit obtained by Sunyaev \& Zel'dovich (1980b).  The relativistic correction becomes significant for large values of $X$, $X>5$.  The distribution of the spectral intensity of the polarized radiation has been also calculated.  The maximum value of $\Delta I_{pol}/y$ is $11\times10^{-6}$ at $X \approx 5$ for $k_{B}T = 10$keV, $\beta_{x} = 1/300$.  At present the observation of the polarization of the CMBR caused by the motion of a cluster of galaxies is not feasible.  However, it is hoped that its observation becomes possible in the future. (See Sazonov \& Sunyaev (1999) for the future projects.)

  We are very grateful to Professor R. Sunyaev, Dr. F. Hansen, and Dr. A. Challinor for their very informative communications.  In particular, we would like to thank Dr. A. Challinor for pointing out an error in the original manuscript of the present paper.  We also wish to thank the referee for many valuable suggestions on the original manuscript which have helped us tremendously in the revision of the paper.  This work is financially supported in part by the Grant-in-Aid of the Japanese Ministry of Education, Science, Sports, and Culture under the contract \#08304026 and \#10640289.


\references{}
\reference{} Acquista, C., \& Anderson, J. L. 1974, ApJ, 191, 567
\reference{} Audit, E., \& Simmons, J. F. L. 1998, preprint astro-ph/9812310 
\reference{} Berestetskii, V. B., Lifshitz, E. M., \& Pitaevskii, L. P. 1982, $Quantum$ $Electrodynamics$ (Oxford: Pergamon)
\reference{} Buchler, J. R., \& Yueh, W. R. 1976, ApJ, 210, 440
\reference{} Challinor, A., Ford, M., \& Lasenby, A. 1999, preprint astro-ph/9905227
\reference{} Challinor, A., \& Lasenby, A. 1998, ApJ, 499, 1
\reference{} Challinor, A., \& Lasenby, A. 1999, ApJ, 510, 930.
\reference{} Hansen, E., \& Lilje, P. B. 1999, preprint astro-ph/9901066
\reference{} Itoh, N., Kohyama, Y., \& Nozawa, S. 1998, ApJ, 502, 7
\reference{} Kompaneets, A. S. 1957, Soviet Physics JETP, 4, 730
\reference{} Landau, L. D., \& Lifshitz, E. M., 1975, $The$ $Classical$ $Theory$ $of$ $Fields$ (Oxford: Pergamon)
\reference{} Nagirner, D. I. 1994, Astronomy Letters, 20, 358
\reference{} Nozawa, S., Itoh, N., \& Kohyama, Y. 1998, ApJ, 508, 17
\reference{} Rephaeli, Y. 1995, ApJ, 445, 33
\reference{} Rephaeli. Y., \& Yankovitch, D. 1997, ApJ, 481, L55
\reference{} Sazonov, S. Y., \& Sunyaev, R. A. 1998, ApJ, 508, 1
\reference{} Sazonov, S. Y., \& Sunyaev, R. A. 1999, preprint astro-ph/9903287
\reference{} Stebbins, A., 1997, preprint astro-ph/9709065
\reference{} Sunyaev, R. A., \& Zel'dovich, Ya. B. 1972, Comm. Ap. Space Phys., 4, 173
\reference{} Sunyaev, R. A., \& Zel'dovich, Ya. B. 1980a, Ann. Rev. Astron. Astrophys., 18, 537
\reference{} Sunyaev, R. A., \& Zel'dovich, Ya. B. 1980b, Mon. Not. R. astro. Soc., 190, 413
\reference{} Sunyaev, R. A., \& Zel'dovich, Ya. B. 1981, Astrophysics and Space Physics Reviews, 1, 1
\reference{} Weymann, R. 1965, Phys. Fluid, 8, 2112
\reference{} Zel'dovich, Ya. B., \& Sunyaev, R. A. 1969, Astrophys. Space Sci., 4, 301


\newpage

\centerline{\bf \large Figure Captions}

\begin{itemize}

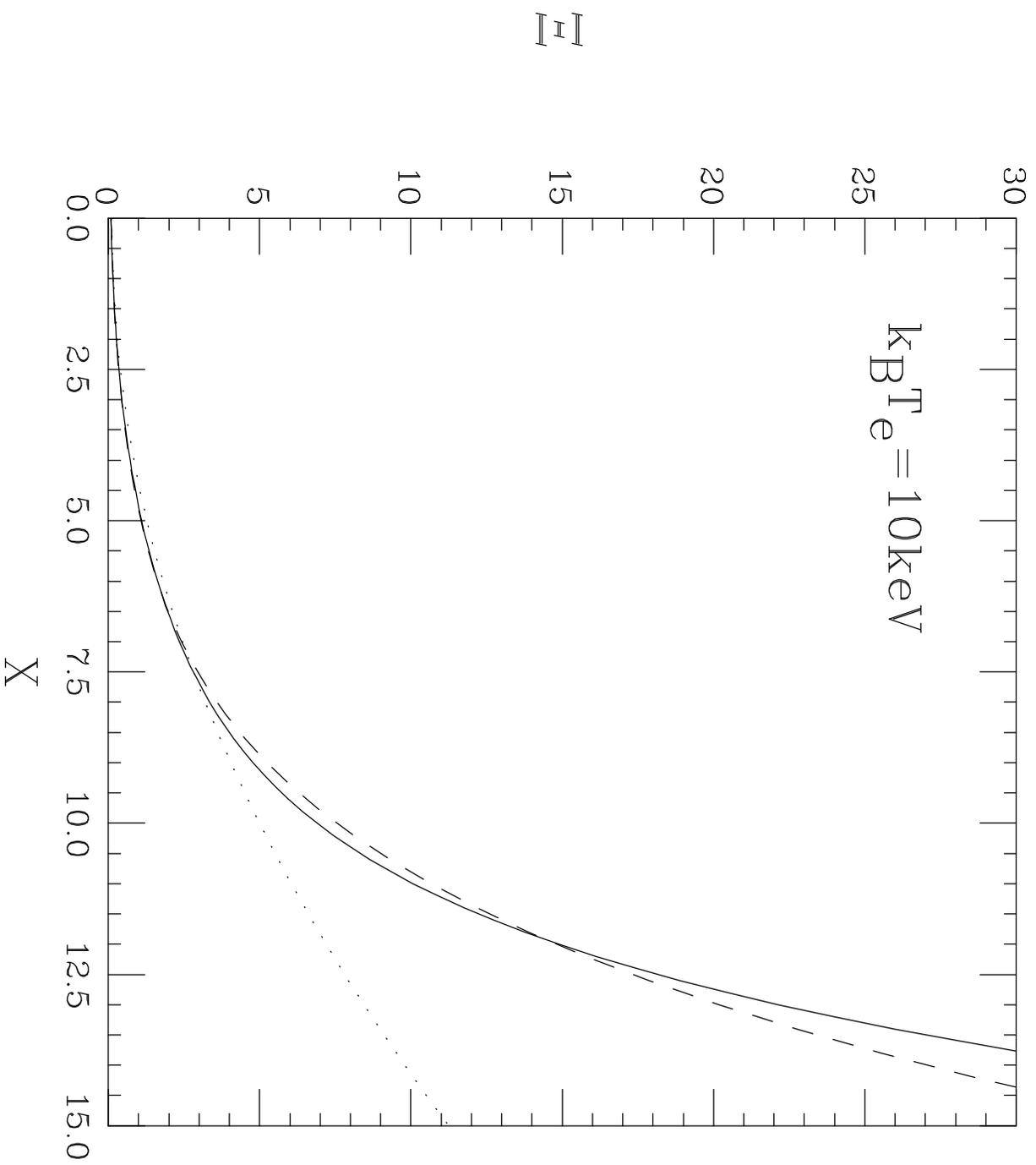
\item Fig.1. Graph of the function $\Xi$ defined by equation (3.16) for $k_{B}T_{e}=10$keV.  The dotted curve is the leading order ($F_{0}$ term) contribution.  The dashed curve is the contribution which includes the $O(\theta_{e})$ term.  The solid curve is the full contribution which includes the $O(\theta_{e}^{2})$ term.

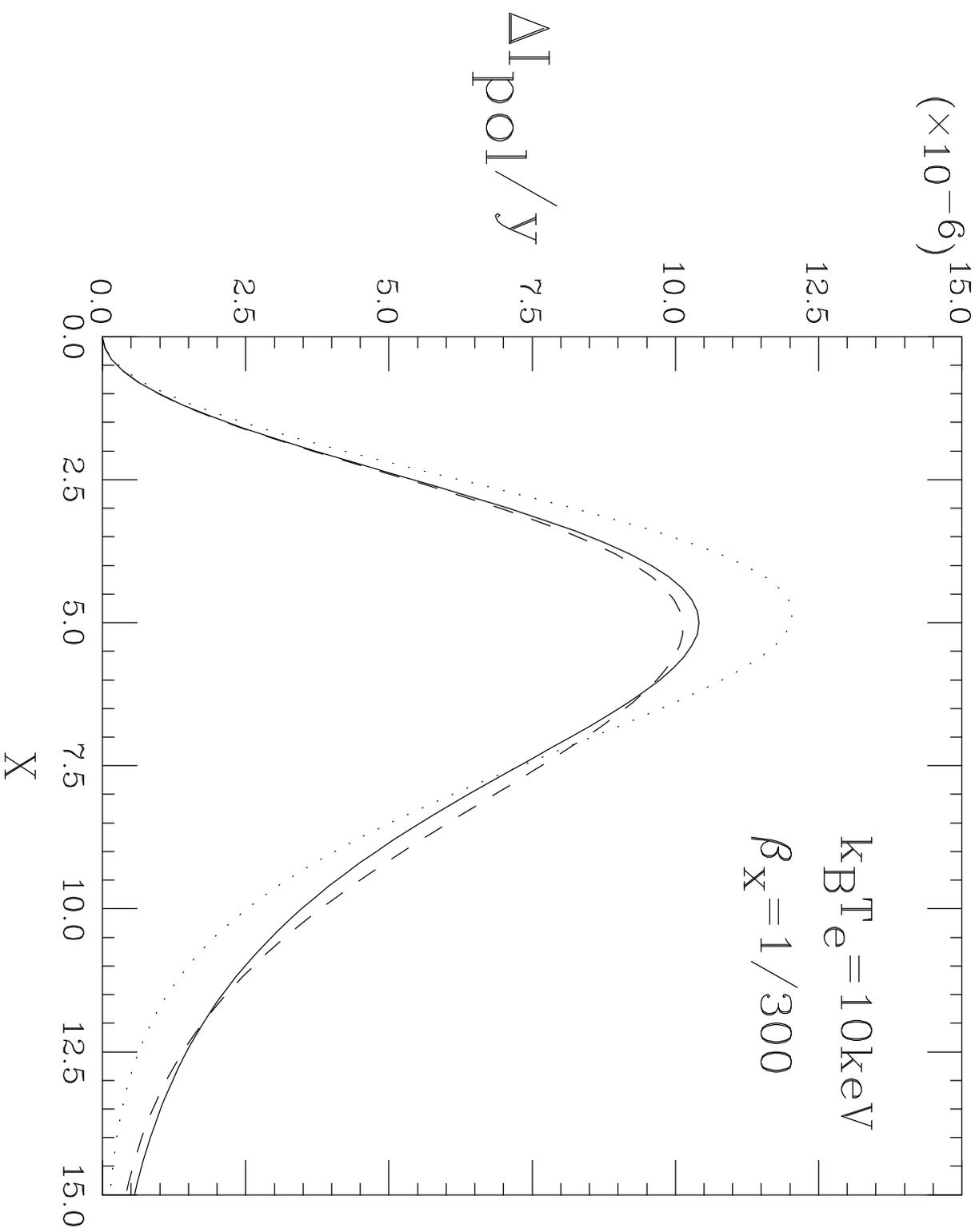
\item Fig.2. Graph of $\Delta I_{pol} /y$ defined by equation (3.17) for the case of $k_{B}T_{e}=10$keV, $\beta_{x}=1/300$.  The dashed curve is the contribution which includes the $O(\theta_{e})$ term.  The solid curve is the full contribution which includes the $O(\theta_{e}^{2})$ term.

\end{itemize}

\end{document}